\documentclass[epjST]{svjour}

\usepackage{graphicx}
\usepackage{amssymb, amsmath}
\begin{document}
\title{Yu-Shiba-Rusinov bound states versus topological edge states in Pb/Si(111) }

\author{
Gerbold C. M\'enard\inst{1,}\inst{4}\fnmsep\thanks{\email{gerbold.menard@nbi.ku.dlk}} \and Christophe Brun\inst{1} \and Rapha\"el Leriche\inst{1} \and Mircea Trif\inst{2,5} \and Fran\c{c}ois Debontridder\inst{1} \and Dominique Demaille\inst{1} \and Dimitri Roditchev \inst{1,}\inst{3} \and Pascal Simon\inst{2,}\thanks{\email{pascal.simon@u-psud.fr}} \and Tristan Cren\inst{1,}\thanks{\email{tristan.cren@upmc.fr}}}
\institute{Institut des Nanosciences de Paris, Universit\'{e} Pierre
et Marie Curie (UPMC), CNRS-UMR 7588, 4 place Jussieu, 75252 Paris, France \and Laboratoire de Physique des Solides, CNRS, Univ. Paris-Sud, Universit\'e Paris-Saclay, 91405 Orsay Cedex, France \and  Laboratoire de physique et d'\'etude des mat\'eriaux, LPEM-UMR8213/CNRS-ESPCI ParisTech-UPMC, 10 rue Vauquelin, 75005 Paris, France \and Center for Quantum Devices and Station Q Copenhagen, Niels Bohr Institute, University of Copenhagen, Universitetsparken 5, 2100 Copenhagen, Denmark \and
Institute for Interdisciplinary Information Sciences, Tsinghua University, Beijing}
\abstract{
There is presently a tremendous activity around the field of topological superconductivity and Majorana fermions. Among the many questions raised, it has become increasingly important to establish the topological or non-topological origin of features associated with Majorana fermions such as zero-bias peaks. Here, we compare in-gap features associated either with isolated magnetic impurities or with magnetic clusters strongly coupled to the atomically thin superconductor Pb/Si(111). We study this system by means of scanning tunneling microscopy and spectroscopy (STM/STS). We take advantage of the fact that the Pb/Si(111) monolayer can exist either in a crystal-ordered phase or in an incommensurate disordered phase to compare the observed spectroscopic features in both phases. This allows us to demonstrate that the strongly resolved in-gap states we found around the magnetic clusters in the disordered phase of Pb have a clear topological origin.
}
\maketitle

\section{Introduction}
\label{intro}

The pursuit of convincing clear signatures of Majorana fermions is at the center of the attention of the condensed matter community. There exists a vivid debate \cite{liu2018,mourik2012,nadjperge2014} about whether the observed zero-bias signatures measured in nanowires emerge from actual topologically protected objects or from trivial Shiba or Andreev bound states \cite{liu2018}.

Scanning tunneling microscopy and spectroscopy (STM/STS) techniques offer one of the most complete approach when it comes to investigate electronic behaviors. This experimental technique allows probing the local density of states at surfaces with an atomic resolution. As any technique, it also has its own inconveniences as it requires a conducting substrate and is restricted to systems with clean surfaces. Hence, many interesting systems studied by mesoscopic transport cannot be accessed by STM. For instance, systems such as semiconducting nanowires proximitized with a superconducting capping, from which possible signatures of Majorana fermions were first observed \cite{mourik2012} are difficult to study using this method. On the other hand, atomically thin superconductors such as Pb/Si(111) \cite{zhang2010,brun2014,brun2017} are perfectly suited since the intrinsic bidimensional aspect of these systems allows probing the whole material.

In this paper we describe and compare two different types of in-gap features observed in Pb/Si(111). The first features are Yu-Shiba-Rusinov (YSR) bound states \cite{yu1965,shiba1968,rusinov1969} induced by individual magnetic atoms. We  discuss the effect of such local perturbation on the 2D superconducting electrons for two different phases of Pb/Si(111) monolayers: the $\sqrt{7}\times\sqrt{3}$ phase and the stripped incommensurate (SIC) phases \cite{zhang2010,brun2014}. We will then focus on a new kind of in-gap edge states we found recently around magnetic Co clusters embedded below the Pb monolayers (first studied in \cite{menard2017}, see also \cite{bjornson2017}).

\section{Yu-Shiba-Rusinov bound states}
\label{ysr_bs}

\begin{figure*}[h!]
\begin{center}
\includegraphics[width = 0.9\textwidth]{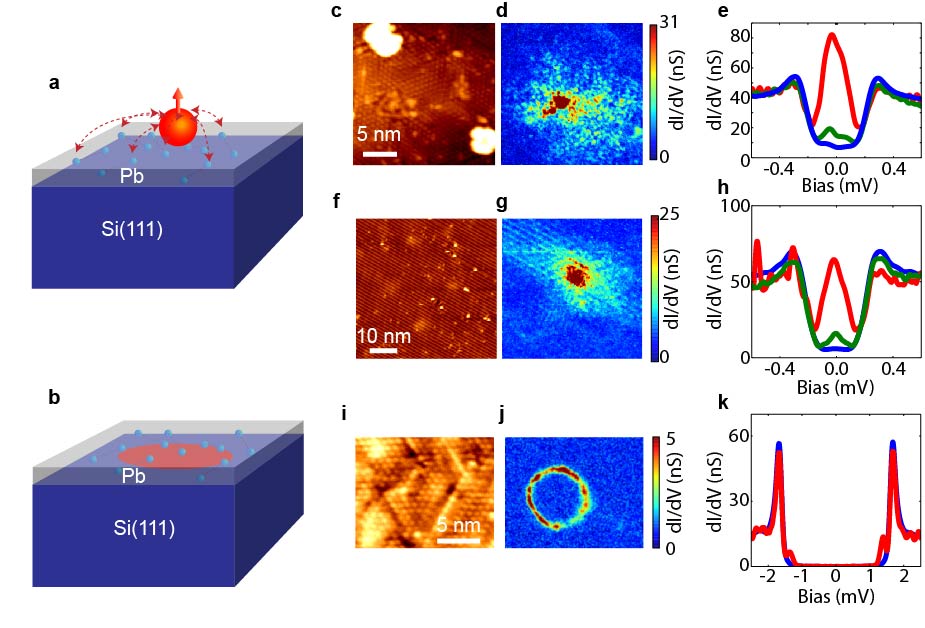}
\caption{\textbf{High-resolution spectroscopy of an individual magnetic impurity in a disordered superconducting Pb/Si(111) monolayer:} \textbf{a} Schematics of the physical situation for an individual magnetic impurity (YSR bound state case). \textbf{b} Schematics of the physical situation for a magnetic cluster (topological region case). \textbf{c} Topography of the $21\times21$ nm$^2$ scanned area with atomic resolution in the SIC phase. \textbf{d}. Conductance map integrated over the superconducting gap showing the spatial dependance of the YSR bound state. \textbf{e}. Selected spectra taken at the largest conductance point (red), 7 nm away (green) and on the side of the measurement area (blue). \textbf{f, g, h} Same as c. d. and e. for the $\sqrt{7}\times \sqrt{3}$ phase of Pb/Si(111). \textbf{i, j, k} same as c. d. and e. in for a buried magnetic cluster. k shows only two spectra, one taken on the edge of the bright ring seen in j. (red) and one taken far away from any impurity on the monolayer (blue). The conductance map j was obtained at a finite bias corresponding to the tip gap 1.41 meV. (colors online)}
\label{fig1}
\end{center}
\end{figure*}

Magnetic impurities in a superconducting substrate give rise to YSR bound states. It has been shown experimentally that the amplitude of the YSR bound states wave function depends on the dimensionality of the substrate \cite{menard2015,ruby16,choi17}. The wave function of YSR bound states is quite complex as it includes an oscillatory term at the scale of the Fermi wavelength convoluted by an envelope term with an exponential decay at the scale of the coherence legnth times an algebraic decay that depends on dimensionality. In two dimensions (2D) the algebraic decay is as $1/\sqrt{r}$ while in 3D it follows a $1/r$ law. As STS is sensitive to the density of states and thus to the square modulus of the wave function, an algebraic decay of the Shiba states in $1/r^2$ is expected in 3D  and in $1/r$ in 2D. This difference  may  seem negligible  but in practice it has some spectacular consequences. In the first STM measurements of YSR states around individual impurities in 3D systems, the wave function was vanishing extremely rapidly (less than 1~nm) with the distance to the impurities \cite{yazdani1997,ji2008} and no spatial oscillations of the YSR states could be measured. We recently analyze the effect of magnetic impurities in 2H-NbSe$_2$, a material with a 2D electronic structure, and were able to visualize the complex spatial structure of the wave function of YSR bound states \cite{menard2015}. A six-fold star shape pattern with branches that extend up to 8 nm from the impurities was observed and electron-hole oscillations with a period given by the Fermi wavelength was observed as predicted by Rusinov in 1969 \cite{rusinov1969}. Since this work, similar complex YSR pattern were also observed in 3D materials such as Pb(111) and Pb(100) in which the spatial extend was found smaller. However, even in 3D materials the YSR states can extend quite far in certain directions due to a kind of channeling effect related to the anisotropy of the Fermi surface: in certain location, the Fermi surface can be 2D-like, which leads to some longer-range Shiba states in these particular directions \cite{ruby16}.

\subsection{Spectroscopic characteristics}
The schematic view of the system we consider is sketched in fig. \ref{fig1}.a.
In Figs \ref{fig1}.c and \ref{fig1}.f we present the topography of two areas belonging respectively to the SIC and $\sqrt{7}\times \sqrt{3}$ phases of Pb/Si(111). The SIC phase appears as an isotropic disordered surface reconstruction while the $\sqrt{7}\times\sqrt{3}$ appears as domains containing stripes that can exist in three different orientations. Fig. \ref{fig1}.f shows one orientational domain. Two other orientations rotated by 60$^\circ$ and 120$^\circ$ are also measured on other locations of the sample. Figs \ref{fig1}.d and \ref{fig1}.g show conductance maps integrated in bias between -0.2 mV and 0.2 mV. For these two maps a central impurity can be seen (high conductance, red color, colors online) from which emerges a real space conductance pattern. For both samples, the impurities are native to the Pb source used for the growth and, though unidentified, is most probably Fe, Co or Cr (possible pollutants in our triple e-beam evaporator used for Pb deposition).

A YSR bound state gives rise to a pair of in-gap states in the superconducting spectra. Such YSR states measured in Pb monolayers are shown on Figs \ref{fig1}.e for the SIC phase and on \ref{fig1}.h for the $\sqrt{7}\times\sqrt{3}$ phase. In both cases we present a selection of three spectra taken on the impurity site (red), 5 nm away from the impurity (green) and far from the impurity (blue). For the SIC spectra shown in Fig. \ref{fig1}.e, we can observe a pair of very intense peaks whose amplitude diminishes as we move away from the center of the impurity. In the case of the $\sqrt{7}\times\sqrt{3}$, a strong peak at zero-bias, with a slight shoulder at positive energy, is observed whose amplitude also decreases when moving away from the center of the impurity. The difference in energy position of these states is easily explained by looking at the equation giving the energy of the YSR bound states \cite{menard2015}. The energy of the YSR states involves both the magnetic exchange interaction and the scattering potential. Therefore, the exact position of the in-gap states depends on the position in the crystal and its chemical nature and may vary a lot between samples. However, one feature that is everlasting is that the energy position of the YSR states we observe in STS do not vary as function to the distance to the impurities. No energy dispersion is measured, one impurity will give rise to one (or a few \cite{ji2008,ruby16}, depending on crystal field effects) peak at a given energy.

\begin{figure*}[h]
\begin{center}
\includegraphics[width = 0.85\textwidth]{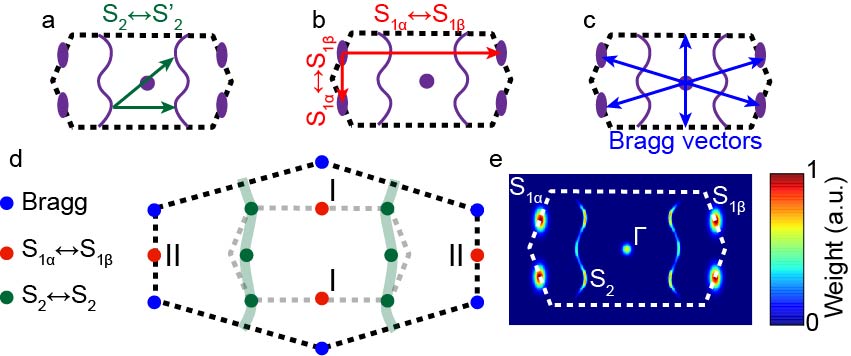}
\caption{\textbf(colors online) {Transitions and Fermi surface for the $\sqrt{7}\times\sqrt{3}$ phase:} \textbf{a} Schematic representation of scattering from S2 pockets onto themselves. \textbf{b}. Schematic representation of transitions from S1 pockets to S1 pockets. c. Schematic representation of transitions from S1 pockets of the same type. \textbf{d}. Diffraction spots in reciprocal space related to the transitions described in a, b and c. The dark dashed lines represent the lines joining the first Bragg points. We kept the first Brillouin zone as a guide for the eye in light grey. \textbf{e}. Model Fermi surface based on ARPES data \cite{kim2010} on this system extrapolated for a single domain. The color code refers to the density of states for each point of the Fermi surface.}
\label{fig2}
\end{center}
\end{figure*}

\subsection{Spatial pattern}
\label{spatial_pattern}

Let us now turn our attention to the spatial structure of the YSR pattern in the conductance maps. For the ordered $\sqrt{7}\times\sqrt{3}$ we see that the YSR bound state spatial pattern follows the symmetry of the underlying crystal and forms lines in the same direction as the atomic raws observed in the topography. This pattern can be directly related to the Fermi surface of the material shown in Fig. \ref{fig2} \cite{kim2010}, the quasiparticle interferences (QPI) occurring along well defined axes described in figs. \ref{fig1}.a-c. The particular scattering channel depicted on Fig. \ref{fig1}.b leads to a YSR pattern with exactly twice the atomic periodicity of the atomic lattice. This expected structure is in perfect agreement with the measurements. 

In stark contrast with this case, the SIC phase exhibits a YSR bound state whose spatial behavior is speckle-like. This finds its origin in the disordered nature of the incommensurate Pb layer which does not benefit from a well defined Fermi surface and thus randomizes the QPI. The only relevant scale in this case is the typical atomic distance that fixes the size of the spots from the speckle pattern. The role of disorder is thus made clear in the context of YSR bound states in which no long range ordered structure are expected to appear in the case of a disordered superconductor except if some underlying mechanism, such as a topological protection, prevents against scattering.

\section{Topological superconductivity}

One of the reason why topology gets so much attention in condensed matter physics is due to the so called topological protection. This means that some kinds of local perturbations of a topological system will leave the system essentially unaltered. In the context of Pb/Si(111) disordered SIC phase, this means that only topologically protected features will be observable as coherent objects at scales for which YSR are observed as speckle pattern, i.e. at scales larger that the interatomic distances.

One way to obtain such protected states is to use a local Zeeman field in order to induce a topological transition in a Rashba superconductor, such as a Pb monolayer. We have recently studied this kind of device and we found that it indeed leads to some clear hint of a topological transition \cite{menard2017}. The system studied is based on the SIC phase described above except that a very small amount of Co was deposited on Si(111) prior to the deposition of the Pb layer. This intermediate step, followed by an annealing at around 375$^\circ$C, leads to a clusturization of the Co atoms into Si-Co domains located just below the Pb monolayer. By this process one get insulating magnetic clusters of sizes ranging from 5 to 40 nm that provokes a strong magnetic exchange field into the superconducting Pb monolayer. The geometry of the samples is presented in Fig. \ref{fig1}.b as a schematic view. The topography of the sample is shown on Fig. \ref{fig1}.i  and the corresponding zero energy conductance map is shown on Fig.\ref{fig1}.j. Note that in order to get a higher energy resolution a superconducting Pb tip was used for this measurement while for the Shiba states discussed above a normal Pt tip was used (in all the cases the measurement were performed at 300~mK). Therefore, the conductance map at zero energy shown in fig.\ref{fig1}.j was measured at a tunnel bias shifted by the tip gap $\Delta_{tip}=1.41$ meV (the precise methodology used for the deconvolution of the tip DOS is detailed in ref. \cite{menard2017}). The most intriguing feature of fig.\ref{fig1}.j is that a very regular edge state is observed, it appears as an almost circular ring of thickness 0.7 nm. This is in striking contrast with the conductance map of YSR bound states around a magnetic impurity observed in the same SIC Pb/Si(111) monolayer (fig. \ref{fig1}.d) which exhibits a speckle like pattern.

\begin{figure}[h]
\begin{center}
\includegraphics[width = 0.55\textwidth]{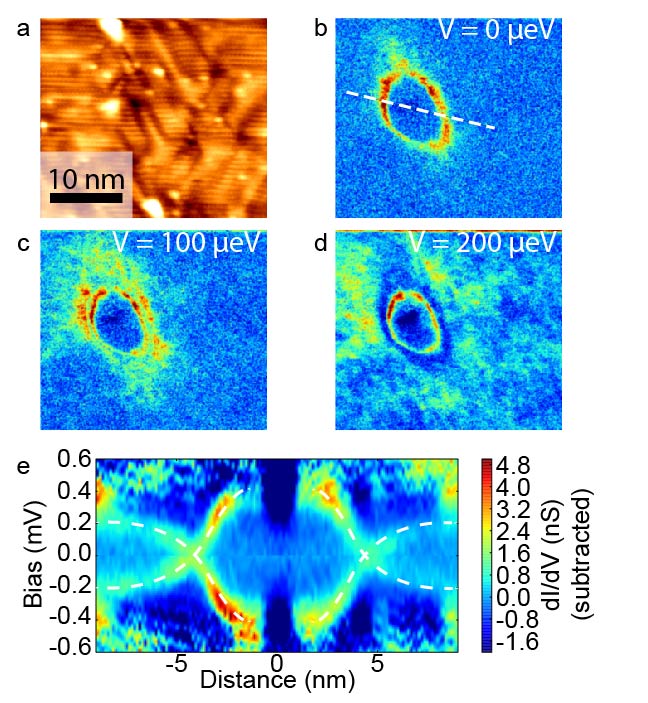}
\caption{\textbf{(colors online) Bias dependence of a topological edge state:} - \textbf{a} Topography of an area of the SIC Pb/Si(111) monolayer including a topological domain. \textbf{b}-\textbf{d} In-gap conductance maps of the same area showing the spatially dispersing topological edge states. The indicated voltages correspond to the corrected voltages corrected from the tip superconducting gap. \textbf{e} Radial cut of the conductance data along the white dashed line from b showing the spatial evolution of the topological edge state.The data was obtained with a superconducting tip. The bare superconducting gap was subtracted to show more clearly the in-gap features.}
\label{fig3}
\end{center}
\end{figure}

In Fig. \ref{fig3} we present yet another example of edge states observed experimentally in a sample different from the one presented in Fig. \ref{fig1}.i. In this configuration we recover the same basic characteristic states crossing the superconducting gap. The underlying Co cluster here is elongated while the one from Fig. \ref{fig1}.i is almost circular. As a result, the edge states we image elongated as well.

\begin{figure}
\begin{center}
\includegraphics[width = 0.9\textwidth]{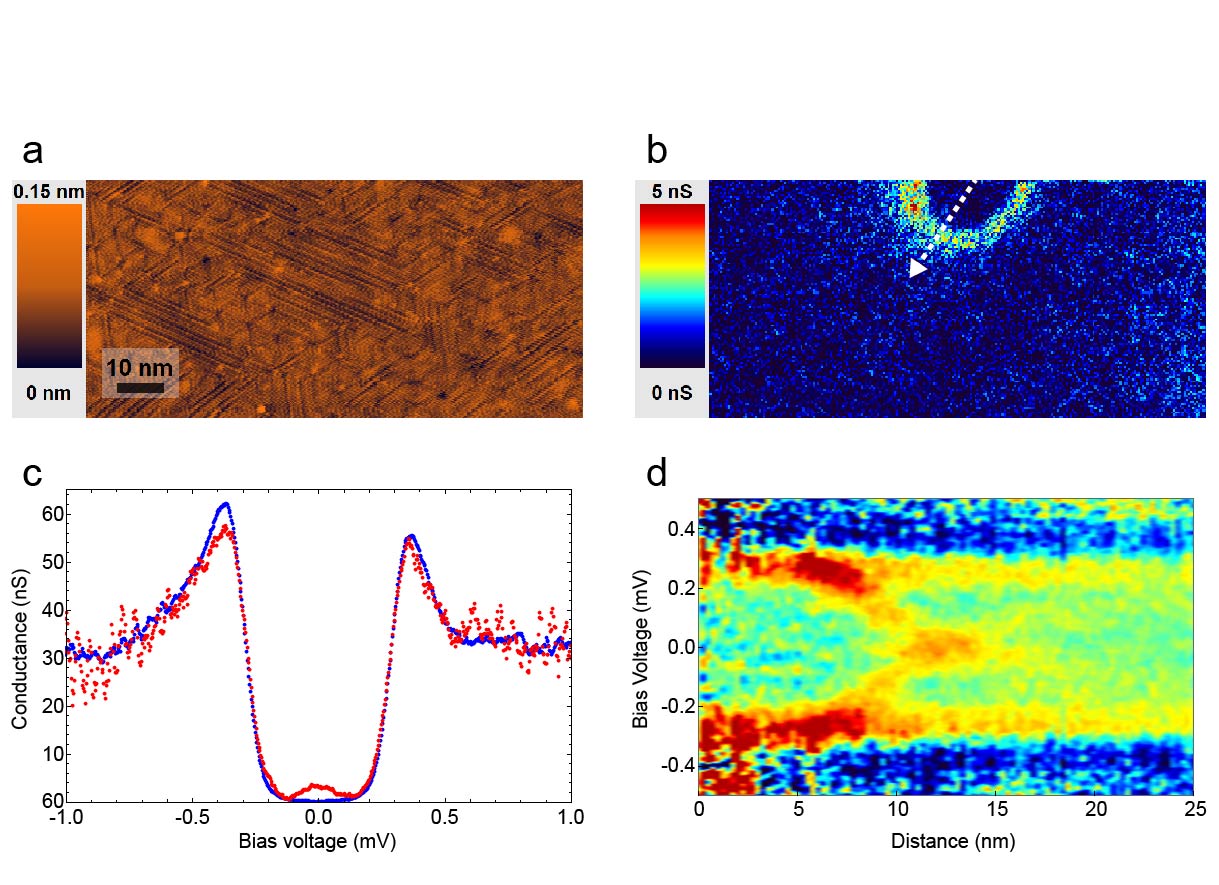}
\caption{\textbf{(colors online) Observation of a topological edge state with a normal tip:}\textbf{a} STM topography of a Pb/Si(111) monolayer with an approximate coverage of 1.26-1.28 monolayer.
\textbf{b} Conductance map at zero bias of the area shown in a, it exhibits a ring-shaped
structures. \textbf{c} dI/dV tunneling spectra measured far from the ring (blue) and on
top of the ring (red). \textbf{d} Line cut from the center of the ring to 25 nm away (following the arrow from b), showing topological edge states dispersing in real space throughout the sample superconducting gap. For clarity a
reference spectrum (blue curve in c) was subtracted in order to enhance the contrast. }
\label{normal_tip_edge}
\end{center}
\end{figure}

In order to put aside any suspicion of tip related artifacts, we present in Fig. \ref{normal_tip_edge} yet another example of this kind of edge state acquired with a normal tip, in the exact same condition as for the YSR bound states we first described in Fig. \ref{fig1}. Once again, the sharp ring is observed with an associated dispersion presented in Fig. \ref{normal_tip_edge}.f.

\section{Theoretical description}
\subsection{Model Hamiltonian}
On the theory level, YSR bound states and topological edge states are both associated with magnetic inhomogeneities, however of different size.
In order to describe qualitatively the observed features, we rely on symmetry to build an effective phenomenological Hamiltonian. Pb/Si(111) breaks inversion symmetry and has therefore a strong Rashba spin orbit coupling. Moreover, most of the electronic spectrum is dominated by the band around the $\rm \Gamma$ point in the Brillouin zone. Combining these two features, we proposed in \cite{menard2017} the  following Boguliubov de-Gennes (BdG) Hamiltonian:
\begin{equation}
H(k) = \xi_k \tau_z +\Delta_S \tau_x+V_z \sigma_z+\left(\alpha\tau_z + \frac{\Delta_T}{k_F}\tau_x\right)(\sigma_x k_y-\sigma_y k_x).
\label{eq1}
\end{equation}
In Eq. (\ref{eq1}), $\xi_k = k^2/2m-\mu$ is the band dispersion near the $\rm\Gamma$ point; $\tau_j$ and $\sigma_j$ (with $j=x, y, z$) are Pauli matrices acting in the particle hole and the spin space respectively. $V_z$ is the Zeeman splitting term and we define $k_F=\sqrt{2m\mu}$. Because inversion symmetry is broken, the superconducting order parameter will also contain a triplet component $\Delta_T$ \cite{rashba} together with the usual singlet component $\Delta_S$. We therefore need to obtain the eigen-energies of 
$H_{Top}=\sum_{k}{\Psi_k^\dagger H(k)\Psi_k}$, 
with the Nambu basis spinor given by $\Psi_k^\dagger=\begin{pmatrix}\hat{c}_{k\uparrow}^\dagger &  \hat{c}_{k\downarrow}^\dagger  & \hat{c}_{-k\downarrow} &  -\hat{c}_{-k\uparrow}
\end{pmatrix}$.

The difference between a Shiba bound state and a topological edge state resides in the definition of $V_z$. A Shiba bound state appears for a very localized potential such as $V_z(r) = |V|\delta(r)$, while a potential extending over distance much larger than the atomic size can lead to a local topological transition \cite{menard2017}. The experimental results shown in figures \ref{fig1}.j and \ref{normal_tip_edge} were obtained with magnetic clusters with diameters of roughly 10~nm, which is much larger than the atomic length scale but smaller than the coherence length $\xi\approx50$~nm.

In order to simplify the discussion let us first consider a spatially invariant configuration (infinite magnetic cluster) to study the topological properties of the Hamiltonian  Eq. (\ref{eq1}) for different configurations of the parameters. In a translationally invariant case (for $V_z(r)=V_z$), the 4$\times$4 matrix defined in Eq. (\ref{eq1}) can easily be diagonalized in order to obtain the eigen-energies of the system. We find four solutions, two electron-like and two hole-like with symmetric energies with respect to the Fermi level. These are given by the following expression
\begin{equation}
E_\pm^2(k)=V_z^2+(\alpha k)^2+\Delta_S^2+\Delta_T^2\frac{k^2}{k_F^2}+\xi_k^2+2\sqrt{V_z^2(\Delta_S^2+\xi_k^2)+\frac{k^2}{k_F^2}(\Delta_S\Delta_T+\alpha k_F\xi_k)^2}.
\label{eq:2}
\end{equation}

\begin{figure}
\begin{center}
\includegraphics[width = 0.9\textwidth]{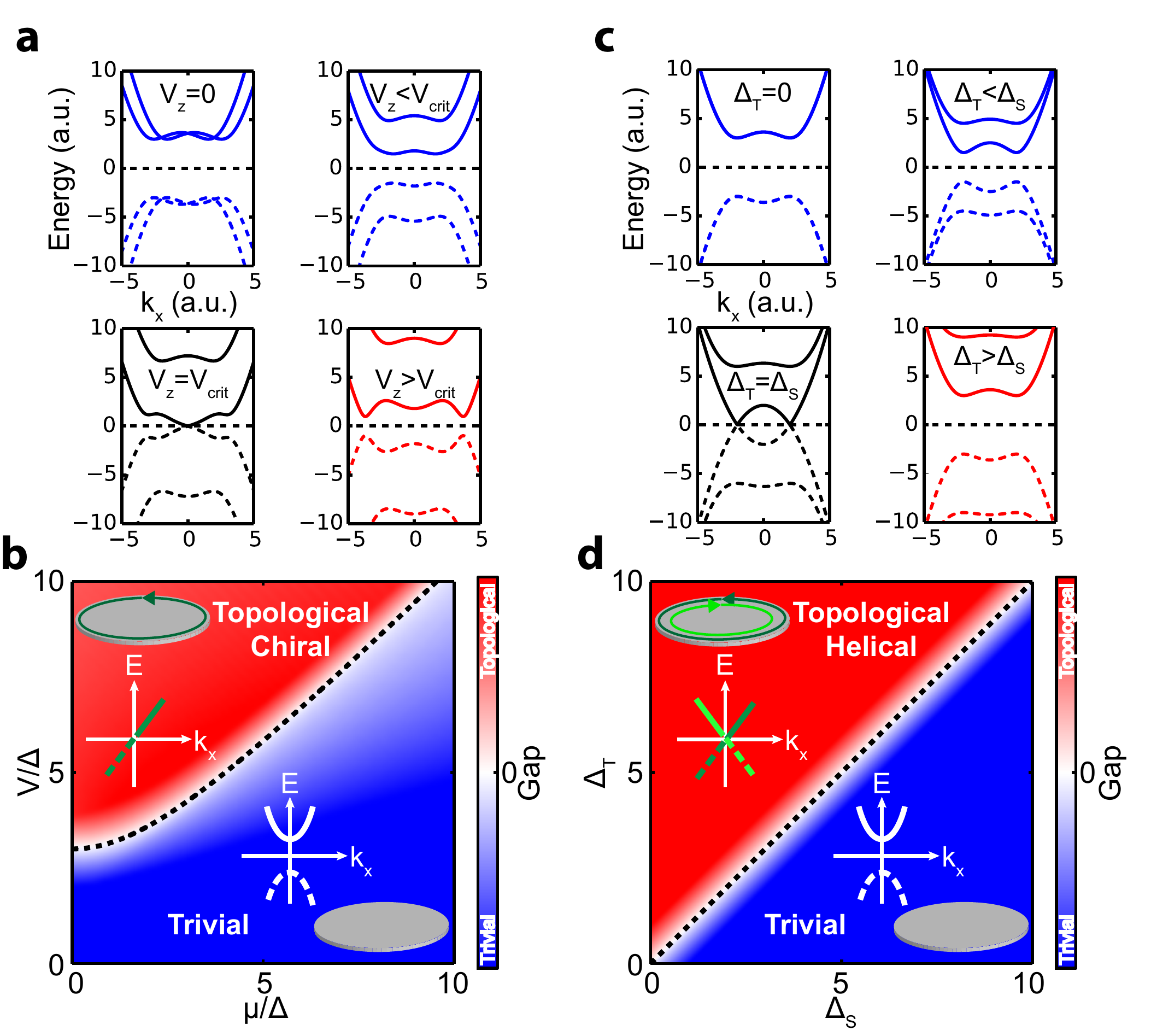}
\caption{\textbf{(colors online) Phase transition diagrams for topological superconductivity} -  \textbf{a} Evolution of the band structure of an s-wave superconductor with Rashba spin-orbit coupling for different values of the Zeeman interaction. The black dashed line indicates the transition between the topological and trivial regimes. \textbf{b} Phase diagram of the system parametrized by the chemical potential $\mu$ and the strength of the Zeeman coupling $V$. \textbf{c} Evolution of the band structure of a superconductor for increasing values of the amplitude of the p-wave component. \textbf{d} Phase diagram of the system parametrized by the amplitude of the singlet ($\Delta_{S}$) and triplet ($\Delta_{T}$) order parameters. The black dashed lines in c and d indicate the transitions between the topological and trivial regimes. }
\label{phase_diag}
\end{center}
\end{figure}

From this dispersion relation it becomes possible to compute the points at which the gap will close (when the system actually undergoes a topological transition) before reopening as a function of the different parameters. Figures \ref{phase_diag}.b and \ref{phase_diag}.d present two phase diagrams calculated in the limits $\Delta_T=0$ and $\alpha=0$ respectively.

In the first case ($\Delta_T=0$), the dispersion in Eq. (\ref{eq:2}) can be simplifed into
\begin{equation}
E_\pm^2(k) = V_z^2+(\alpha k)^2+\Delta_S^2+\xi_k^2\pm 2\sqrt{V_z^2(\Delta_S^2+\xi_k^2)+(\alpha k\xi_k)^2},
\label{eq:3}
\end{equation}
from which we can observe a topological transition from a trivial superconductor to a chiral one at a critical field $V_{z,crit}=\sqrt{\Delta_S^2+\mu^2}$ where the gap closes at $k=0$ (see Fig. \ref{phase_diag}.a). It is worth noting here that even though the spin-orbit coupling does not explicitly intervene in the expression of the critical field, its presence is absolutely necessary to guarantee the reopening of the gap at larger fields. It is easily seen that for $\alpha = 0$, the gap does not reopen and thus prevents the appearance of a gapped topological superconducting phase.

The second case shown in Fig. \ref{phase_diag} corresponds to the condition $\alpha = 0$ and $V_z=0$ for which the superconductor is time-reversal invariant, and we expect a transition from a trivial to a helical state. In this case, the topological transition is controlled by the amplitude of the triplet term $\Delta_T$. From Eq. (\ref{eq1}) we can write in that case
\begin{equation}
E_{\pm}^2(k)=\left(\Delta_S\pm\frac{k}{k_F}\Delta_T \right)+\xi_k^2.
\label{eq:4}
\end{equation}
The topological transition is thus obtained for $\Delta_S=\Delta_T$ at $k=k_F$ and is presented in Fig. \ref{phase_diag}.d.

By combining both a triplet order parameter and a Zeeman field, one surprisingly finds that the superconducting helical phase  survives at finite $V_z$ despite time-reversal symmetry being broken. There is whole
 transition line between the trivial and the helical phase. This transition line is found from Eq. (\ref{eq:2}) as $\Delta_T=\sqrt{\Delta_S^2-V_z^2}$ with the gap closing happening in this case at $k=k_F$. 

\subsection{Chern number}

The bulk spectrum along with the gap closing points (or lines) do not suffice to account for the number of edge states in a system with boundaries occurring in each phase. For that, we calculate the Chern number for each phase (the symmetry being chiral, a Chern number fully describes the number of edge states). Topological and trivial superconductor are ultimately discriminated by their Chern number. A Chern number $C=0$ will describe a trivial superconductor while a non-zero Chern number will describe a topological superconductor. The Chern number is defined as the integral curvature of the filled band $n(k)$, which reads
\begin{equation}
C= \sum_{n}{\frac{1}{2i\pi}\int_{BZ}{d^2kF_n(k)}},
\label{eq:5}
\end{equation}
where $F_n(k)=\frac{\partial A_{n,y}(k)}{\partial{x}}- \frac{\partial A_{n,x}(k)}{\partial{y}}$ and $A_{n,\mu=x,y}(k)=\langle n(k)|\partial_{\mu}|n(k) \rangle$ are  the Berry flux and Berry connection for the band $n$ respectively. These can be evaluated by calculating the eigenstates of the Hamiltonian  in Eq. (\ref{eq1}). Using the numerical method exposed in \cite{fukui2005} we find the Chern number to be non-zero only inside the chiral region, \textit{i.e.} for $V_z>\sqrt{\Delta_S^2+\mu^2}$, independently of whether or not we include spin-orbit coupling.
In order to get a more detailed glimpse at the topology of the system in the helical case, it is instructive to analyze a simple limiting case $\Delta_S = 0$ and $\alpha = 0$ for which the Hamiltonian is spin block diagonal
\begin{equation}
H = \begin{pmatrix}
H_+ & 0 \\
0 & H_-
\end{pmatrix},
\end{equation}
where $H_\pm = (\xi_k\pm V_z)\sigma_z +\frac{\Delta_T}{k_F}(\sigma_x k_y \mp \sigma_y k_x)$.

We can then calculate the Chern number corresponding to each block. We find $C_\pm = \pm1$, which gives one pair of (quasi-) helical states, or two chiral states at the edges of the sample. Note that the total Chern number of the system is $C=C_+ +C_- = 0$, as found from the full Hamiltonian. As time-reversal invariance is broken, note that these states are not conjugate Kramers pairs. This in turn implies that the edge states can separate spatially, as found in experiment. The analysis of a more complete Hamiltonian (considering finite $\Delta_S$ and $\alpha$) does not alter these results.

\subsection{Real space topological transition}
We have analyzed the homogeneous model Hamiltonian in Eq. (\ref{eq1}). We can use these results to interpret what could happens in our experiment. Far away from the cluster, the substrate is in  the trivial phase. Above the cluster, we can assume that the Zeeman exchange field drives the system into a chiral topological phase. Therefore, there must be some gap closing and associated dispersive edge states around the perimeter of the magnetic cluster to connect the trivial and topological superconducting phases as the ones shown in Fig. \ref{fig3}.

\section{Conclusion}

In this paper, we have discussed the experimental signatures of two types of perturbations to the superconducting condensate in Pb/Si(111). We first showed that the structure of the monolayer influences the spatial pattern of the YSR bound states in such a way that disordered phases present speckle-like pattern while ordered phases reflect the structure of the electronic transition from the Fermi surface. The SIC monolayer that exhibits a speckle like pattern for YSR states around atomic impurities shows a radically different response to extended magnetic domains. Some regular edge states apparently insensitive to the local disorder are observed. These edge states present an energy dispersion that is associated to a kind of splitting in real space. This offers a strong argument in favor of the topological interpretation in the context of a topological protection of these states. In order to reinforce this interpretation it would be interesting to put a vortex in the middle of the magnetic domains in order to check if a zero energy Majorana bound state is found as expected for a vortex core in a topological superconductor.
xz
\section{Author contributions}
G.C.M., C.B., R.T.L. and T.C carried out the STM/STS experiments and D.D. Performed the TEM measurements. G.C.M. and T.C. processed and analyzed the data. P.S., G.C.M., M.T. and T.C. performed the theoretical modeling. G.C.M, P.S. and T.C. wrote the manuscript. D.R., T.C. and F.D. designed the experimental setup. All authors discussed the results presented in the manuscript.

\end{document}